\documentclass[reprint,10pt]{revtex4-1}

\usepackage{colortbl}
\usepackage{floatrow}
\usepackage{graphicx}
\definecolor{light-gray}{gray}{0.95}

\begin{document}

\title{Multiplex lexical networks reveal patterns in early word acquisition in children}

\author{Massimo Stella}
\affiliation{Institute for Complex Systems Simulation, University of Southampton, UK}
\email{Corresponding author: massimo.stella@inbox.com}
\author{Nicole M. Beckage}
\affiliation{Department of Electrical Engineering and Computer Science, University of Kansas, USA}
\author{Markus Brede}
\affiliation{Institute for Complex Systems Simulation, University of Southampton, UK }

\begin{abstract} 

Network models of language have provided a way of linking cognitive processes to language structure. However, current approaches focus only on one linguistic relationship at a time, missing the complex multi-relational nature of language. In this work, we overcome this limitation by modelling the mental lexicon of English-speaking toddlers as a multiplex lexical network, i.e. a multi-layered network where N=529 words/nodes are connected according to four relationships: (i) free association, (ii) feature sharing, (iii) co-occurrence, and (iv) phonological similarity. We investigate the topology of the resulting multiplex and then proceed to evaluate single layers and the full multiplex structure on their ability to predict empirically observed age of acquisition data of English speaking toddlers. We find that the multiplex topology is an important proxy of the cognitive processes of acquisition, capable of capturing emergent lexicon structure. In fact, we show that the multiplex structure is fundamentally more powerful than individual layers in predicting the ordering with which words are acquired. Furthermore, multiplex analysis allows for a quantification of distinct phases of lexical acquisition in early learners: while initially all the multiplex layers contribute to word learning, after about month 23 free associations take the lead in driving word acquisition.

\end{abstract} 

\maketitle

Language consists of a multi-level mapping of meanings onto words \cite{icancho2003,aitchison2012}. In order to
communicate, humans must learn how to use linguistic structures to express thoughts as words. The cognitive processes behind language learning may organise the components of language into a so-called \textit{mental lexicon} (ML) \cite{aitchison2012}. This lexicon can be
described as a network structure of interacting lexical items (e.g. word
representations). Empirical studies in psycholinguistics suggest that, rather than providing
exact word definitions (as in common dictionaries), the ML stores word
meanings as multi-relational or multiplex word patterns \cite{quillian1967word,
collins1969retrieval, collins1975spreading, borge2010, baronchelli2013, dautriche2015learning}. How
the multiplex organisation of the ML relates to and influences language
learning is still poorly understood but new techniques related to multiplex networks \cite{wasserman1994social,kivela2014review,dedomenico2013math,boccaletti2014review} allow us to explore patterns of word acquisition within the mental lexicon. We achieve this by constructing an edge-coloured multiplex network 
\cite{wasserman1994social} based on relational
features of phonology, semantics, and syntax. Going beyond the
topological description of the multiplex network, we exploit the multiplex
structure to predict normative acquisition of young children.

Previous literature on modelling language learning and use through network
science has largely focused on single-layer representations of networks \cite{borge2010,baronchelli2013,beckage2015cognition}. We build on these works, that strongly suggest that many cognitive constraints and mechanisms affecting the use of language can be explored through network science. In fact, experimental results have shown strong correlations between network structure and human performance in
various language related tasks. Measurements of retrieval times
\cite{collins1969retrieval, collins1975spreading, dedeyne2008network}, age
of acquisition  \cite{stey2005large, hills2009longitudinal, beckage2011},
creativity \cite{kenett2015investigating} and
even semantic degradation due to ageing \cite{goni2011semantic} have been
studied and modeled using single layer networks. Further, networks of phonological word similarities have highlighted an upper bound on the size of phonological neighbourhoods as well as a tendency to avoid local clustering \cite{vitevitch2008,stella2015phonological,stella2016investigating}.
Experiments with adults have indicated that these constraints relate to word confusability in identification tasks \cite{vitevitch2008,vitevitch2012}. 

While single layer networks reveal aspects of the structure of language and language-related cognitive processes, it is clear that this approach cannot offer a unified view of language that simultaneously accounts for phonological, semantic and syntactic aspects of language,
as required by increasingly sophisticated experimental set ups 
\cite{dautriche2015learning}. One approach that is capable of overcoming this limitation is to use multiplex network representations as originally introduced in the social sciences \cite{wasserman1994social}. Within this approach, we project the complexity of an individual's mental lexicon (ML) onto a multiplex network, which we call \emph{Multiplex Lexical Network} (MLN). Our MLN is composed of multiple network layers, with nodes representing words and layers capturing different relationships between words. Multiplex networks \cite{wasserman1994social} are a specific type of multi-layer networks \cite{kivela2014review,boccaletti2014review} where nodes represent the same set of items on all layers. In our MLN we do not consider inter-layer connections. Instead we focus on intra-layer relationships that have previously been shown to influence children's lexical learning in single-layer network studies: 1) free-associations
\cite{nelson2004university,stey2005large}, 2) shared  features
\cite{mcrae2005semantic,hills2009longitudinal}, 3) co-occurrence in child
directed speech \cite{macwhinney2000childes,beckage2011} and 4) phonological
similarity \cite{vitevitch2008,wiethan2014early}. We quantitatively show that by understanding the structure of multiple layers of the ML we can gain further insight into human cognition as related to language acquisition in young toddlers. 

The main advantage of the multiplex approach is that it allows for a more
detailed representation and quantification of many real-world systems  \cite{de2016physics}. This added model complexity provides additional insights into a system's structure and
dynamics. In the last few years multiplex modelling has provided novel
insights in areas as diverse as social balance in on-line platforms
\cite{szell2010multirelational,gomez2012evolution}, emergence and stability of multiculturalism \cite{battiston2015interplay}, congestion in
transportation networks \cite{de2016physics}, and ecosystems in ecology
\cite{pilosof2015ecological,stella2016parasite}, cf. 
\cite{kivela2014review,boccaletti2014review} for a review on multi-layer and multiplex networks.

The multiplex network approach relates to other works in psycholinguistics investigating language via multi-layer networks. Liu et al. \cite{liu2014empirical} analysed Chinese as a multi-layer network, composed of syntactic and phonemic layers. They found that almost half of the syntactic dependency relations are between phonologically similar words. A similar analysis was performed for English and Croatian establishing language specific relationships between syntax and phonology \cite{martinvcic2016multilayer}. However, these studies consider only topological features of a given language representation. We extend this approach by explicitly considering the interaction of topological features with a cognitive dynamical process, i.e. for exploring language acquisition in young children.

Below, we consider the ability of structural network features to account for normative acquisition trajectories. We evaluate performance using empirical parent reports of toddlers' productive speech, aggregated to capture the order of word learning for children between 16 and 30 months \cite{Dale1996}. Our aim is to assess the predictive power of a multiplex lexicon representation to capture emerging features and possible mechanisms of word acquisition in young children. As it currently stands, our model is solely descriptive: it only approximates importance of words from observational data rather than inferring it from a generative Bayesian framework \cite{xu2007word}. However, we demonstrate that our multiplex model is more accurate than any single layer representation in predicting the normative acquisition trajectories of young children. We find that the multiplex network can account for developmental trends, offering predictions and interpretability that analysis based on single layer networks or word specific measures such as frequency or length cannot. 

\section*{Results} 
\subsection*{Multiplex network construction and analysis}  

\begin{figure*}[ht]
\centering
\includegraphics[width=17cm]{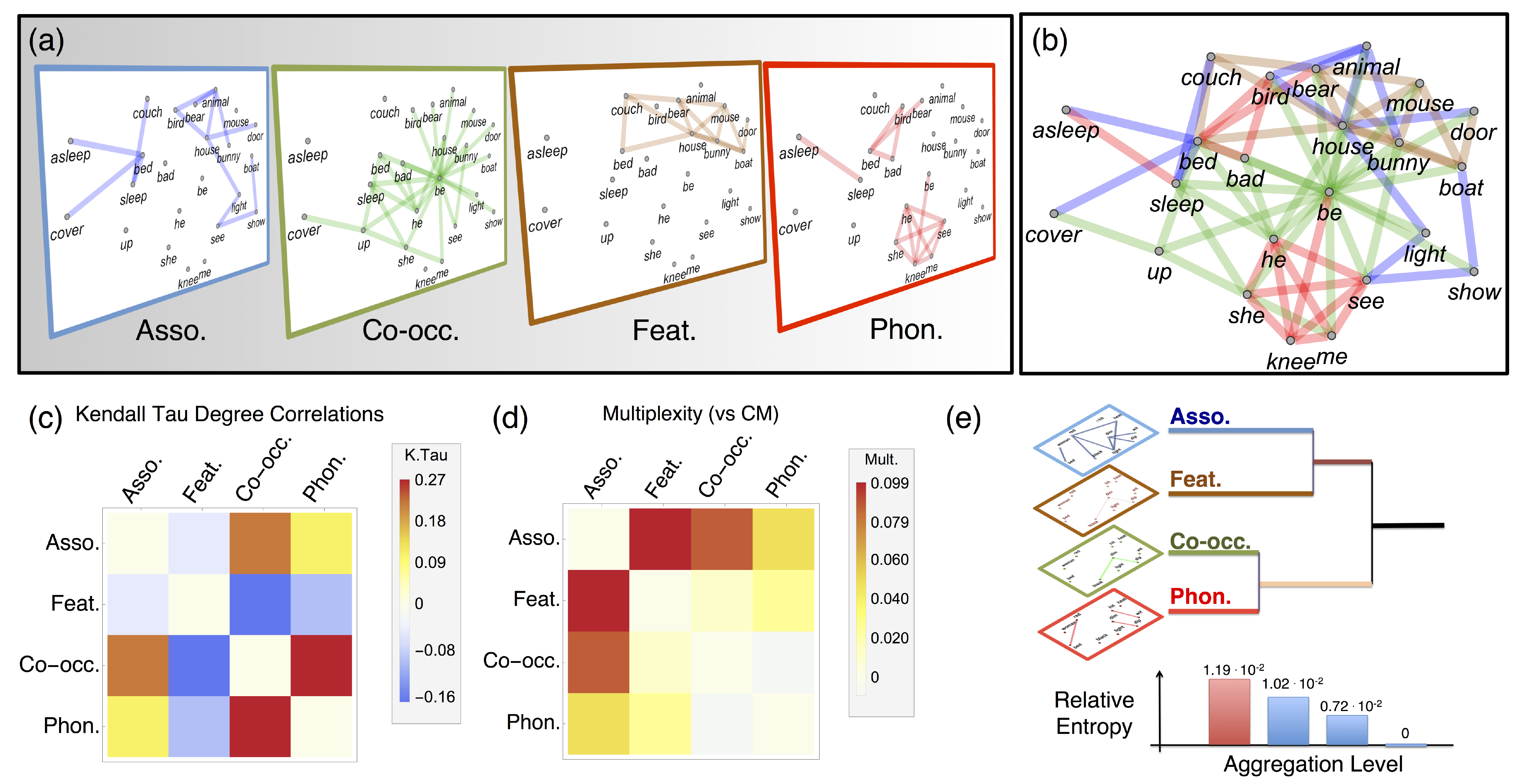}
\caption{(a) Visualisation of part of the multiplex network representing a toddler's lexicon. (b) Visualisation of an example of the MLN as an edge coloured network. Both (a) and (b) include only 24 of the 529 nodes/words in the whole MLN. (c) Degree correlations across layers quantified by the Kendall Tau. (d) Multiplexity or edge overlap among different layers relative to expectations from configuration models. (e) Reducibility dendrogram for the MLN. The MLN is irreducible: its layers are so different from each other that aggregating any of them would lead to a loss of information about topological patterns.} 
\label{fig:visua}
\end{figure*} 

We first construct a multiplex lexical network composed of four layers capturing i) free associations from the South Florida association norms \cite{nelson2004university}, (ii) feature sharing from the McRae et al. dataset \cite{mcrae2005semantic}, (iii) co-occurrence in child-directed speech from  the CHILDES dataset \cite{macwhinney2000childes}, and (iv) phonological similarities from WordNet 3.0 \cite{vitevitch2008} (cf. Methods and SI Sect. 2). Hence the resulting MLN properties emerge from the assembly of independent measures rather than from a multiplex network design. Figures~\ref{fig:visua} (a) and (b)  provide a visualisation of the links in each of the four MLN layers. Panel (a) treats each layer as separate whereas panel (b) describes the multiplex as an edge-coloured graph \cite{wasserman1994social}, in which different colours denote different relationships. As evident from (a) and (b), even if words might be disconnected on one layer, they could be connected on the  MLN structure. 

We report single-layer summary statistics and comparison to configuration models (i.e. random graphs preserving the degrees of an empirical network \cite{newman2010}) in Table~\ref{tab:1}. We focus on network features already analysed in previous works \cite{motter2002,sigman2002,stey2005large,vitevitch2008}: degree, clustering, degree mixing, connectivity ratio and mean shortest path length. For instance, the degree of a node counts its connections \cite{newman2010} and on the phonological layer, it coincides with the size of its phonological neighbourhood \cite{vitevitch2008,vitevitch2012}. See Methods and SI Sect. S3 for precise definitions and interpretations. 

Tab.~\ref{tab:1} shows that the phonological layer has a smaller mean degree compared to other layers. Furthermore, apart from the association layer all other layers are highly fragmented. Their largest connected components include between 24.2\% and 62.4\% of the 529 words. These percentages quantify how many words could be reached from each other by navigating through links on individual layers. In contrast, the \textit{multiplex network} is connected \cite{dedomenico2014navigability} -- every pair of words is connected by a path that potentially exploits different relationship types. Compared to configuration models, except for the co-occurrence layer all other layers display higher clustering, non-zero assortativity and slightly larger path length in their largest connected component. In agreement with previous work \cite{gravino2012complex,sigman2002,vitevitch2014,beckage2015cognition} these patterns suggest that the MLN layers display a core-periphery structure. From a cognitive perspective this feature might possibly facilitate navigation from word to word within cores (see SI Sect. S3 for further discussion).  Layers also differ in their degree distributions, see SI Fig. S2: they are exponential-like for the synonym and phonological layers but much more heavy-tailed for the co-occurrence and association layers. Heavy tails indicate the presence of network hubs, i.e. words which are significantly more connected than average. These hub words play a central role for navigation through concept space \cite{sigman2002,motter2002,gravino2012complex} and could therefore be good candidates for earlier acquisition.

Is it appropriate to represent the ML in terms of four separate layers or
can we get the same type of topological information with fewer layers? Structural reducibility analysis \cite{de2015structural} can test if multiplex layers can be aggregated without losing information (see SI for details). Results reported in Fig.~\ref{fig:visua} (e) show that aggregation cannot be performed without information loss, demonstrating that the chosen 4-layer representation is in fact \textit{irreducible} \cite{de2015structural}. This is not to say that each layer plays a role in language acquisition, just that the information encoded in each layer is different from all other layers. 

Before turning to modeling word acquisition, we investigate the MLN structure in more detail. Specifically, Fig.~\ref{fig:visua} (c) analyses the similarity between degrees of a word on different layers. We find that words in the feature and co-occurrence layers tend to have negative degree correlations (Kendall Tau $\kappa \approx -0.16, p<0.0001$), indicating that hubs in one layer tend to have lower degrees in the other layers. This may suggest a type of semantic differentiation \cite{stey2005large} in English in which words that have similar features tend to not be used in close proximity of each other, at least when speaking to
young children. In contrast, the co-occurrence and phonological layers display positive degree correlations (Kendall Tau $\kappa \approx 0.27, p<0.0001$): in the children's lexicon, words having many co-occurrences also have larger phonological neighbourhoods, further supporting the idea that phonological similarities and co-occurrences in child directed speech
influence each other \cite{macwhinney2000childes,wiethan2014early}. 

Another question of interest is to what extent links overlap across layers. This can be quantified by edge overlap (or multiplexity
\cite{gemmetto2015multiplexity}). Results are reported in
Fig.~\ref{fig:visua} (d). As expected, we find some overlap between
connections in the feature, co-occurrence and association layers; these layers capture semantic aspects of the ML. Beyond the overlap between these layers,
links tend not to overlap significantly more than one would expect in configuration model across the other layers. This suggests that different MLN layers tend to capture
different relational aspects of the ML, again emphasising that the MLN is a more complete model than single network representations. We now explore the importance of individual layers and of the MLN  as a whole in explaining acquisition trajectories.

\begin{table*} 
\footnotesize
\caption{Metrics for the MLN layers with $N=529$ nodes, listing mean degree $\left\langle k \right\rangle$, mean clustering coefficient $CC$, assortativity coefficient $a$, percentage of nodes in the largest connected component $Conn.$, and mean shortest path length of the largest connected component $\left\langle
d\right\rangle$. The quantities are measured for the layers of the MLN and for randomized equivalents with the same degree sequences (denoted with CM). Error bars represent standard deviations and are reported in brackets behind the last significant digit.} 
\label{tab:1} 
\begin{tabular}{p{3cm}p{0.5cm}p{0.5cm}p{0.8cm}p{0.8cm}p{0.8cm}|p{0.9cm}p{1.1cm}p{1.4cm}p{1.1cm}}
\hline\noalign{\smallskip}
Empirical Network & $\left\langle k \right\rangle$ & $CC$ & $a$ & $Conn.$ & $\left\langle d\right\rangle$ & $CC\,(CM)$ & $a\,(CM)$ & \textit{Conn.} (CM) & $\left\langle d\right\rangle \,(CM)$\\
\hline\noalign{\smallskip}
\rule{0pt}{1em}Associations (Asso.) & 9.3  & 0.20  & -0.1 & 99.6\% & 3.2 &  0.03(1) &  $-0.01(6)$ &  99.6\% & 3.00(1)\\
\rule{0pt}{1em}Feature Norms (Feat.)    & 9.0 & 0.63 & -0.01 & 24.2\% & 1.8 &  0.38(2) & $-0.07(6)$  &  24.2\% & 1.72(1) \\
\rule{0pt}{1em}Co-occurrences (Co-occ.) & 8.1 & 0.69 & -0.44 & 62.4\% & 2.2 & 0.75(3)&  $-0.40(2)$ & 62.4\% & 2.19(1) \\
\rule{0pt}{1em}Phonological Sim (Phon.) & 1.31 & 0.37  &  0.48 & 33.1\% & 7.7 & 0.02(1) & $0.03(2)$ & 45.5\%(8) & 5.31(7) \\
\rule{0pt}{1em}Multiplex Aggregate      & 26.5 & 0.33 & -0.07 & 100\% & 2.4 & 0.18(5) &  $-0.10(8)$ & 100\% & 2.25(1) \\
\noalign{\smallskip}\hline\noalign{\smallskip}
\end{tabular}
\end{table*} 

\subsection*{Multiplex Orderings: Results and Discussion} 

Here we explore the process of word learning through the MLN assuming a
preferential acquisition scenario \cite{hills2009longitudinal},
where words are learned earlier if they are central in the
language environment. We assume that relevance of the language environment
can be summarised by connectivity in a language network \cite{sigman2002,motter2002,baronchelli2013,beckage2015cognition} rather than making assumptions on the statistics of relationships to infer relevance, as is more standard in a Bayesian approach \cite{xu2007word,frank2009using}.

We generate a word acquisition ordering $\tau$ by ranking words according to measures computed on the MLN. We assume that if one ranking is predictive, it would indicate that the feature generating it is relevant to early language learning. We compare MLN-based orderings to those based on non-relational, word-specific information such as word frequency and word length. For consistency with the literature, we investigated features previously considered in linguistic networks such as word degree \cite{stey2005large,beckage2011,vitevitch2014}, closeness \cite{motter2002,sigman2002}, betweenness \cite{borge2010,motter2002,sigman2002} and PageRank \cite{griffiths2007google}. For more details see Methods and SI Sect. S6.

We compare our orderings against an ensemble of empirical age of acquisition orderings, obtained from a probabilistic interpretation of the normative age of acquisition of words based on the MacArthur-Bates Communicative Development Inventory\cite{macwhinney2000childes} (CDI) and over a population of roughly 1000 toddlers (see Methods). The comparison is performed by computing the average word overlap $O(\tau,t)$ of ordering $\tau$ with normative orderings, namely how many words $\tau$ correctly predicts as learned until the inventory includes $t$ words. Word \textit{gains} are then obtained by subtracting $O(\tau,t)$ from the expected number of correctly predicted words by random guessing (see Methods and SI Sect. S6). Word gains are normalised by the inventory size $t$ and their Z-scores relative to statistics from random guesses are estimated. Positive \textit{vocabulary normalised word gain} indicates better than random performance of ordering $\tau$ and negative values indicate performance inferior to random guessing.

\begin{figure*}[ht] 
\centering
\includegraphics[height=4.9cm]{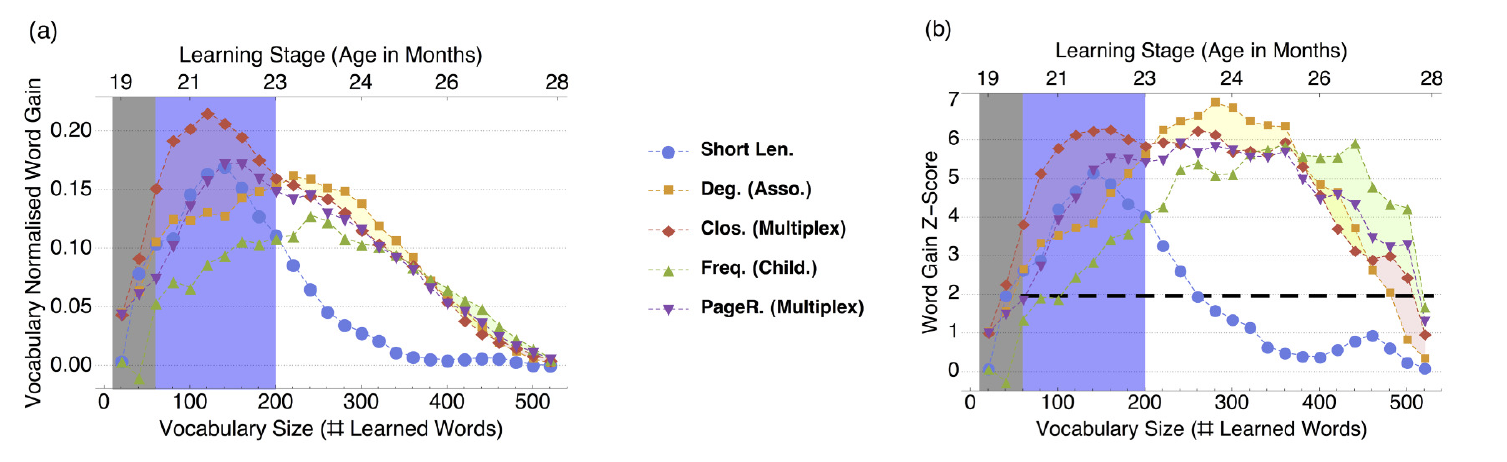}
\caption{Left: Vocabulary normalised word gains for different orderings: shorter words first (Short Len.), words with higher degree in the association layer first (Deg. Asso.), closer words on the whole MLN first (Clos. Multiplex), more frequent words first (Freq. Child.) and higher multiplex Page Rank first (PageR. Multiplex). A normalised word gain of $0.20$ means that 20\% of words in the vocabulary have been predicted by the ordering strategy on top of expectations from random guessing. Higher word gains indicate better predictability. Right: Statistical significance indicated by word gain Z-score of the respective orderings. The distribution of the random overlaps can be approximated by a Gaussian after 60 words have been acquired (see SI Sect. S5). In this range, represented by the black dashed line, a score $Z\geq 1.96$ denotes word acquisition patterns that are different from random fluctuations with a 2.5\% significance level. 
In both plots, error bars are the same size as the dots. Best performing orderings are
highlighted with different overlays, which identify different learning
stages: VELS (in grey), ELS (in light purple) and LLS (in white). Direct sampling was used in VELS for testing the statistical significance of results.}
\label{fig:orde}
\end{figure*}

Fig.~\ref{fig:orde} reports normalised word gains (a) and their Z-scores (b) for the best performing orderings from our analysis. No ordering always outperforms the others: we notice shifts in which word features are the most relevant for different periods of word learning. This agrees with the idea that the progression of acquisition changes over the course of development: children learn some initial words and then tend to generalise those words based on network connectivity later, i.e. preferential acquisition \cite{hills2009longitudinal}. This cognitive interpretation of the observed shifts at different acquisition stages allow us to distinguish at least three distinct learning phases. In a first stage, comprising months 19 and 20, which we call the \textit{very early learning stage} (VELS), we are trying to predict the first 40 learned words (grey overlays in Fig.~\ref{fig:orde}). Direct sampling \cite{grimmett2001probability} for this period indicates that our results are below the 2.5\% threshold for statistical significance -- the network appears to contain little information relevant for predicting acquisition. 
This phase is followed by an
\textit{early learning stage} (ELS) covering an age range between 20 and 23 months. This phase is characterised by \textit{closeness centrality outmatching} every other ordering (see also SI Sect. S6). Last we discriminate a \textit{late learning stage} (LLS) comprising ages between 23 and 28 months. During this stage the dominant hub words have already been learned and more localized structure is likely affecting learning. LLS is characterised by degrees in the association layer and frequency performing equally to or slightly better than closeness and other word features (cf. SI Sect. S6).

Interestingly, while previous work has found frequency to be a good predictor for normative age of acquisition \cite{kuperman2012age}, results in Fig.~\ref{fig:orde} suggest that word length and some topological network information are more predictive of early acquisition ordering. Good performance is achieved by word length in ELS, possibly because of a \textit{least effort effect} \cite{icancho2003} that suggests that short words are easier to memorise and learn. This hypothesis is also supported by the observation that word length loses its predictive
power at later learning stages (word gain Z-scores are compatible with random fluctuations after 300 words have been learned). We interpret this happens when more sophisticated learning
strategies are being used. One of them might be mediated by degree orderings. Among the four layers, ordering words according to their degree in the association layer gives the best performance and it can be up to 100\% more predictive than random guessing (cf. SI Fig. S7). This confirms previous results suggesting that association norms are generally good predictive models for early language learning \cite{hills2009longitudinal,stey2005large,beckage2015modeling}. We
conjecture associations perform well in our case because they are strongly related with semantic memory
\cite{nelson2004university,zock2015words,dedeyne2008network}.

Previous work has pointed out that in early learning, during VELS and ELS, it is very difficult to outperform random guessing because children may not have a clear strategy for learning new words early in language acquisition \cite{beckage2015modeling}. Instead here we find that better than random prediction is possible even in very early learning stages, at least for normative acquisition. Using the multiplex closeness of words it is possible to predict up to 160\% words more than at random (cf. SI Fig. S7) and up to 25\% more than by using degrees in the association layer (cf. SI Fig. S8). As reported in SI Sect. S6, we checked that multiplex closeness consistently outperforms every single-layer orderings during early learning stages. This is a promising result in that multiplex approaches may allow us to quantify learning strategies in toddlers. Also other multiplex measures such as versatile PageRank \cite{dedomenico2015ranking} perform up to 10\% better than degrees in the association layer (cf. SI Fig. S8). The enhanced prediction ability of multiplex features indicates that the multiplex representation of the ML captures meaningful influences of multiple types of feature representations. Altogether, we can conclude that early learning is strongly related to the multiplex structure, specifically the closeness of words considering \emph{all} layers that form the MLN.

Results for the LLS show a reversal of trends. Here we find
that a single layer based on association degree ordering gives best predictability. Global and local multiplex measures, as well as word specific features such as word frequency, perform similarly or slightly worse until 400 words have been learned (see Fig~\ref{fig:orde} and SI Fig. S8). This indicates that preferential acquisition is very important, but at this point in development the dominant
relevant relationships are those captured by the association layer. This is
compatible with the observed emergence of semantic learning in children
\cite{wiethan2014early}. However it has to be noticed that statistical significance deteriorates markedly during later stages of learning  making comparisons of performance in the LLS difficult.

So far, we evaluated orderings based on single network layers and the combined
multiplex, showing that prediction performance can benefit from taking
account of information from multiple layers. In this analysis we assumed that each layer plays an equal role. In the next section we evaluate the changing
influence of layers during the evolution of the lexicon. 

\subsection*{Optimisation of layer influences} 

\begin{figure*}[ht!] 
\centering

\includegraphics[width=14.8cm]{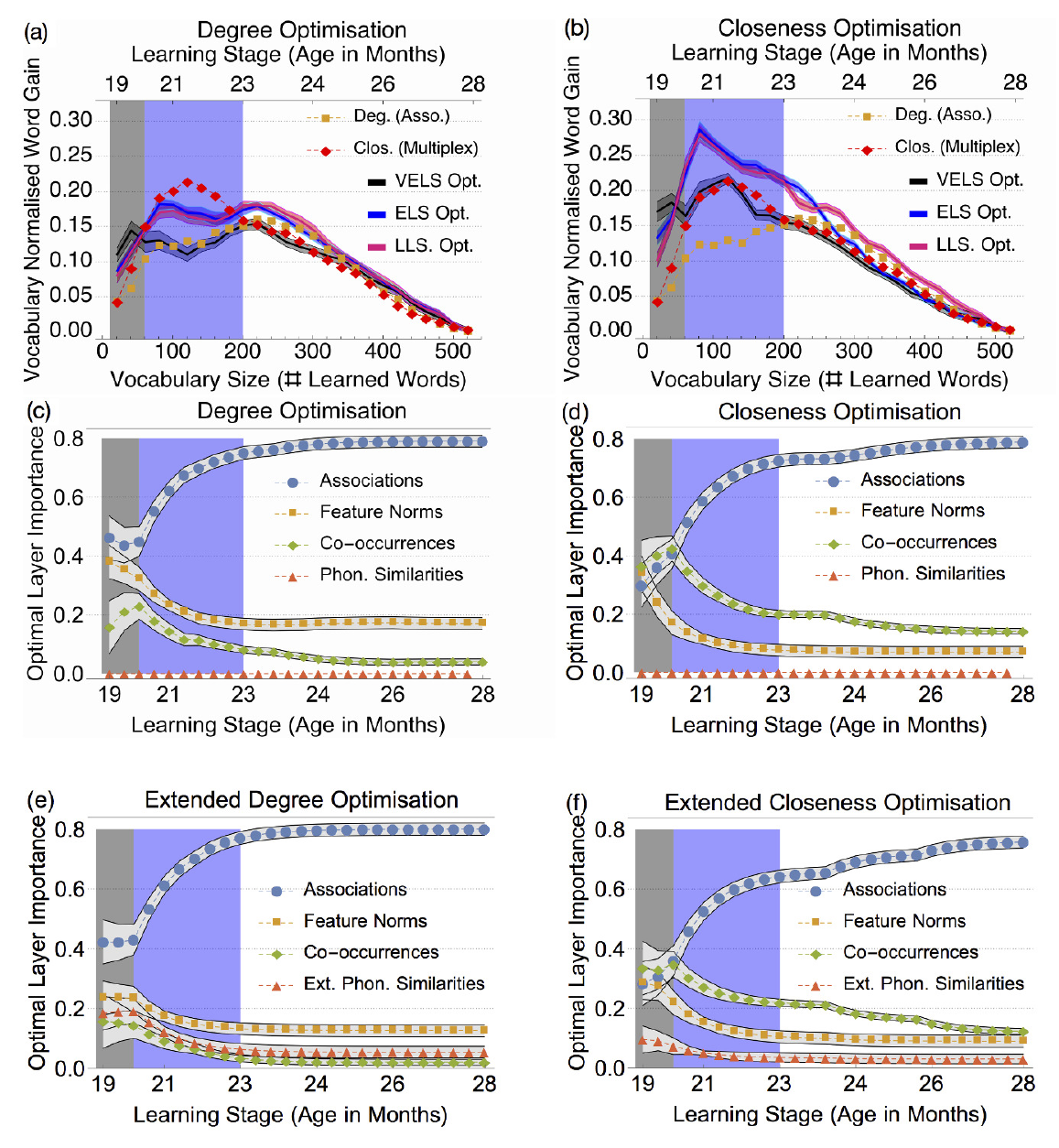}
\caption{Top (a-d): Optimisation results for the original MLN. (a,b): Vocabulary normalised word gains corresponding to the optimal layer influences at the end of VELS (black), middle of ELS (blue), and middle of LLS (pink) for degree optimisation (a) and closeness optimisation (b). Error margins represent standard deviations over randomised ensembles. In (b) no optimisation of degree outperforms multiplex closeness during ELS. (c,d): Average optimal layer weights indicating the influence or importance of layers over different learning stages obtained from Monte Carlo experiments with degree (c) and closeness (d) optimisation. The phonological layer resulted in influences around $10^-3$. The same optimisation trends were obtained when the phonological layer was excluded from the optimisation. Bottom (e,f): For comparison, optimal layer weights for the MLN where rather than the phonological layer with 529 words an extended one with 30,000 words is used, instead. Considering the extended phonological layer leads to word gains at least 6\% better than (a) and (b) (cf. SI Table 3).}
\label{fig:dego}
\end{figure*} 

To explore the influence of different layers on word acquisition, we consider linear combinations of word features on each layer in order to obtain weighted multiplex measures (cf. Methods section). For a given network metric, we optimise the coefficients of the linear combinations to maximise vocabulary normalised word gains. The resulting layer weights, that maximise predictability, indicate the influence of the respective layers on word acquisition. We explored several types of optimisation, based on degree, closeness, betweenness and local clustering (cf. SI Sect. S7 and SI Tab. S3). We focus on the two best performing quantities from the ordering experiments, namely degree and closeness. 

In order to avoid over-fitting, we perform a Monte Carlo robustness analysis: we optimise over subsets of word trajectories sampled uniformly at random, consisting of  only 80\% of the original words. Averages are computed over these trajectories and over different realisations of the age of acquisition trajectories from the normative age of acquisition ordering. We also confirmed that the improvement in performance of optimised multiplex parameters is strongly dependent on the structure and overlap of layers, as experiments on randomised multiplex networks result in far smaller word gains (see SI Sect. S8).

Optimisation results (cf. Fig. \ref{fig:dego}) indicated negligible contribution of the phonological layer, with weights ($\approx ~ 10^{-3}$). Degree and closeness orderings based only on the phonological layer perform poorly as well, yielding word gains compatible with random fluctuations at a 5\% significance level (see SI Sect. S6.1 and SI Fig. S5). We retrieved the same optimisation results for the other layers when the phonological layer was excluded. In addition, we considered phonological word scores coming from an "extended" phonological layer for adults, already analysed by Stella and Brede \cite{stella2015phonological} and including 30000 words (see also SI Sect. S6.1.1). Assessing the influence of adults' phonology over word acquisition in toddlers matches previous approaches in the relevant literature \cite{carlson2014children} and we present results in our MLN framework in Fig. \ref{fig:dego} but also in SI Sect. S7. 

Optimal layer influences are displayed as a function of age for the MLN layers in Fig. \ref{fig:dego} (c) and (d). Similar patterns of layer influence are found for rankings based on both degree and closeness. We see a clear distinction between different learning stages. When degree is considered, cf. Fig. \ref{fig:dego} (c), in VELS the main layer
contributions stem from the association and the feature norm layers. Instead, when optimising closeness, cf. Fig. \ref{fig:dego} (d), in VELS all layers are found to contribute equally. In both cases the ELS, the next learning stage, marks  a transition region between the VELS (before) and the LLS (after) in which only the association layer dominates. This transition is what motivated our choice of distinguishing between VELS and ELS. For degree-based optimisations, contributions from the co-occurrence layer become very small in the LLS, while associations contribute 80\% of weight. An analogous transition is observed in the ELS phase when closeness is optimised and after that associations make up 80\% of the total layer influence. The dominant influence of associations in later stages of learning is indicative of the emergence of preferential acquisition \cite{hills2009longitudinal} that we already observed in ordering experiments. 

When extended phonology is considered, in Fig. \ref{fig:dego} (e) and (f), the influences of semantic layers remain consistent, however, phonology has a higher impact when phonological neighbourhood size/degree is considered rather than closeness. In VELS the extended phonology has importances similar to co-occurrences but the influence of both approaches 0 later in development.

In Fig. \ref{fig:dego} (a) and (b) we compare vocabulary normalised word gains of trajectories based on optimal layer influences against our ordering experiments (cf. SI Fig. S9 for the respective Z-scores and previous section). On the one hand, when predicting only words learned in VELS, the optimal influence orderings outmatch both degree in the association layer and multiplex closeness during VELS but perform worse at later stages. Direct sampling indicates that these results are significant with a p-value of $0.01$. Optimal influences up to the middle of ELS and LLS provide similar word gains, which are both up to 25\% higher than the optimal word gains from VELS. This suggests that the \textit{same learning strategies} may be adopted \textit{after} VELS, starting in ELS and lasting throughout LLS. On the other hand, results in Fig.~\ref{fig:orde} (a) highlight the
importance of closeness centrality for predicting word acquisition at early
stages. In fact, in ELS even the optimal linear combination of degrees results in \textit{smaller word gains} compared to the unweighted multiplex closeness centrality. Optimal linear combinations of closeness perform even better than the unweighted multiplex closeness, see Fig.~\ref{fig:orde} (b). Notice that closeness is a \textit{global} network feature, accounting for the position of a node relative to all other nodes. This is distinct from \textit{local} features, such as degree, which only measure a node's relationship to its first neighbours. Our findings give support to the hypothesis that, particularly in the early learning stages, word learning is influenced by the global structure of the multiplex lexical network. This global multiplex structure is able to capture some important word patterns that influence early lexicon development.

\section*{Conclusions and future work} 

Introducing the framework of \textit{multiplex lexical networks} (MLN), we have described and analysed normative word acquisition patterns of children between the ages of 19 and 28 months. Our approach considers semantic, phonological and syntactic contexts and captures more linguistic information than separate analysis of any single layer network previously could \cite{beckage2011,beckage2015modeling,dedeyne2008network,hills2009longitudinal}. We go beyond a purely topological description by exploiting network
properties to predict the order of normative word learning in children. 
Interestingly, we find that the best topological feature in predicting word learning changes through the course of development. This allows us to use network information to distinguish three stages of learning: (i) a very early learning stage where all but the phonological layer contribute substantially to prediction, (ii) an early
learning stage which marks a transition period, and (iii) a late learning
stage in which contribution from word associations dominates word learning. We believe the last point is due to free associations serving as a much better proxy for detecting superordinate words compared to other network features (cf. SI Sect. 6.3 for a quantitative discussion). Further assessment of the extent to which the taxonomic organization serves as a mediator variable influencing word prediction is an interesting direction for future work.

Comparing the predictive power of various MLN features over time
confirms: (i) the superiority of some multiplex network characteristics
relative to single layer networks and word specific measures such as frequency or word length, and (ii) the special role closeness centrality might play in early word learning. This is emphasised in two ways: (i) strong performance of multiplex closeness, outperforming all combinations of
local network features in the early learning phase, and (ii) the best overall predictions of the order words are learned is related to a weighted multiplex version of single-layer closeness centralities. We thus find strong indications that, particularly at early stages of learning, word acquisition in children is  driven by minimising relational distances of learned words relative to other words in both the semantic and syntactic space of the mental lexicon. 

With the MLN, we explored questions about normative lexical acquisition,
uncovering developmental learning stages and quantifying the influence of
certain types of linguistic information on early acquisition. Nevertheless, our MLN is limited in important ways. It is important to bear in mind that the MLN
representation is only a projection of an individual child's full ML. Additionally, the results presented in this paper consider normative lexical acquisition. That is to say, a specific child may not learn words in the same order as the normative orderings. Instead normative order is obtained by averaging over roughly 1000 productive vocabulary reports. We attempt to address this shortcoming by sampling multiple orders probabilistically from these norms. Extending the model and testing the universality of the learning strategies in VELS and ELS on longitudinal data for individual children would be an interesting future research direction \cite{aitchison2012}. 

Another limitation is the use of a relatively small vocabulary. While the CDI is a commonly used check list of words a child produces, there are many words that a child may know which are not on the CDI. By expanding our model to the full vocabulary of a child, and for longer periods of development, we can further increase our understanding of word learning.

While the present study focuses on word learning in English, results should be corroborated by considering word learning in different languages. Future work can extend our analysis by considering correlations in word learning across other cognitive and linguistic domains to build an even richer picture of language organisation and learning than presented here. A stream of work in psycholinguistics has explored detailed mechanisms of word learning \cite{xu2007word}. Such work typically explores how words are memorised when clues are presented in different order and focuses on the underlying mechanisms of inferring meaning-object mappings, pointing out that learning is strongly dependent on context. Integrating context-dependent information in our MLN framework would surely be an exciting future research direction for investigating individual children's learning trajectories. 

The marginal influence of the original phonological layer found in this study can be reconciled with previous findings. Dautriche \textit{et al.}\cite{dautriche2015learning} showed that 18 months old toddlers failed to learn an object label when it was a phonological neighbour of a noun they already knew. In our dataset 80\% of words learned by toddlers within the first two months are nouns. This noun-richness is in opposition with orderings extracted from phonological degree and closeness, which identify mainly non-noun words, verbs in particular, as learned initially. This trend is not surprising, as having learned a verb earlier makes it easier for toddlers to learn its phonological neighbours \cite{dautriche2015learning}. However, this contrast in noun-richness at early development stages reflects in the poor performance of the original phonological layer in predicting normative acquisition ordering. Nonetheless, we show that phonology influences word acquisition by considering a phonological network extended to the adults' lexicon, like in \cite{carlson2014children}. This extended layer identifies nouns as early candidates for learning and provides significant prediction results. This interplay between phonology and acquisition is in agreement with previous findings in the relevant literature \cite{carlson2014children,stamer2012,vitevitch2012,vitevitch2014}.

We have shown strong evidence that multiplex lexical networks
capture a richer picture of the mental lexicon than previous works using only single-layer networks, in spite of the above limitations. More importantly, this work provides novel formalised methods for exploring and explaining the influence of linguistic features on early acquisition.


\section*{Methods} 
\subsubsection*{Age of acquisition dataset} 

Age of acquisition orderings are constructed based on the MacArthur-Bates
Communicative Development Inventory (CDI) norms \cite{Dale1996}. The CDI,
based on parent report of the productive vocabulary of children aged between
16 and 30 months, has been shown to be related to future language ability
in young children \cite{Dale1996,Fenson1994}. The norms are averaged over
more than 1000 vocabulary reports and indicate the percentage of children,
at a given age, that reportedly produce (and understand) a specific word.
From these norms, we sample orderings of word learning. For example, "mommy" and "ball"
are reported as produced by 93\% and 64\% of children by 16 months, respectively; these words are (usually) learned earlier than "chair" which is produced by only
14\% of 16 month olds. We assume a word is known once 50\% of children in a
given month produce that word. We then sample an order of words learned within a month assuming that higher rates of production probabilistically indicate earlier learning. Starting from position one of the ranking, we: (i) sample a learned word with probability proportional to the percentage of children which are able to produce it, conditioned on the overall production rate being over 50\%, (ii) erase it from the list of candidate words to be learned at later positions and (iii) proceed one position down in the ranking and start from (i). The sampling starts from month 16 and it stops when every word has been sampled. Note that a non-probabilistic version of this ordering has been used in other developmental modelling contexts \cite{hills2009longitudinal,beckage2015modeling}.

\subsubsection*{Construction of the multiplex network} 

The four layers of the MLN were selected because of their use in previous single-layer
network studies investigating word learning of young children
\cite{beckage2011,beckage2015cognition,hills2009longitudinal,borge2010}. The
four layers we consider here are (i) a semantic layer based on the Florida
Free Association Norms \cite{nelson2004university} where a link exists from
word A to word B if word B is a free associate of the cue word A; (ii) a layer capturing feature similarity based on the McRae Feature Norms \cite{mcrae2005semantic} where
words A and B are connected if they share at least one semantic feature;
(iii) a layer based on word co-occurrences (measured in child-directed
speech \cite{macwhinney2000childes}) where words A and B are connected if
they co-occur more than 45 times, where the threshold 45 was chosen
approximately to match the link density of the other semantic layers; and a
(iv) layer capturing phonological similarities (based on IPA transcription
from  WordNet 3.0 \cite{miller1995wordnet}) where words A and B are connected if they have
IPA transcriptions with edit distance one. We treat the association
layer as undirected and edge weights in layers (i) to (iii) are ignored. The
resulting MLN is composed of 529 words which represent the intersection
between the CDI data and the set of words which have at
least one connection on any layer (see SI). 

\subsubsection*{Cognitive interpretation of network metrics} 

We investigate the same local and global features of words/nodes that were
investigated in previous works about network representations of
language: degree, closeness, betweenness and PageRank (see SI). In a given network, the degree of a node is the number of its links \cite{newman2010}. In our MLN, a word is characterised by one degree measure per layer. For instance, on the association layer a word with degree $k^{(Asso.)}$ has $k^{(Asso.)}$ associates. On the phonological layer the degree of a word coincides with its phonological neighbourhood size \cite{vitevitch2014}: a word with degree $k^{(Phon.)}$ has $k^{(Phon.)}$ similar sounding words. Higher degree words in semantic networks were found to have lower age of
acquisition norms \cite{stey2005large}, while degree on the phonological
layer correlated positively with word confusability \cite{vitevitch2012}. On the whole MLN, we can associate each word to its multidegree, namely the sum of its degrees on all the layers \cite{dedomenico2013math}. Fundamentally, the usefulness of measures such as degree would suggest that words with multiple associates, multiple phonological neighbors and multiple
shared features are those words that are learned earliest. The mechanisms
behind this phenomenon could be that words with many associates and shared
features are those words that are important in the language environment
because they play a central role or because they are important to the
parent or child. We remain agnostic as to the exact reason behind the
increased `importance' of a word, but suggest that it can be summarized by the degree. 
Closeness relates to how fast information spreads from a node to others \cite{newman2010}. It is interesting for cognitive science since previous work showed that words closer on semantic topologies tend to be processed together in shorter time
\cite{collins1969retrieval,collins1975spreading,motter2002}. Betweenness captures the extent to which a node falls on the shortest path between pairs of words \cite{newman2010}. It captures centrality of words through a hypothetical navigation on the ML \cite{sigman2002,motter2002,borge2010} (cf. SI Sect. S6). PageRank identifies the likelihood of reaching a given node by a random walk in a network. High PageRank words on semantic networks were found easier to retrieve in fluency tasks \cite{griffiths2007google}. Closeness, betweenness and PageRank on multiplex networks exploit jumps through layers (cf. SI Sect. S6).

\subsubsection*{Overlap measures and word ranking} 

A word trajectory is an ordered list $\tau=(w_1,w_2, ...,w_N)$ indicating
the exact order in which words are learned. We define ${\tau_{aoa}}$ to be the ensemble of 
rankings derived from the CDI norms. We
sample each specific ordering $\tau_{aoa}$ based on the probability of production as reported in the age of acquisition norms.

Predicted word orderings are generated as follows. First, each word is
ranked according to a word score, $s_i$, which is derived from network
specific or extrinsic word features (e.g. frequency or word length). Words are
then ordered according to word score, starting with the largest score. Words
receive a position in the predicted trajectory according to their position
in the ordering of word scores. Ties in word orderings $s_i$ are taken into
account by averaging over all resolutions. For example, one might wish to
consider orderings based on degree in the association layer and would obtain
the following word scores $s_{food}=62$, $s_{water}=45$, $s_{eat}=20$. The 
corresponding predicted acquisition trajectory would be
$\tau=(food,water,eat)$. Apart from the features mention in the main text of
the paper, we also considered many others (cf. SI Sect. 6).

To evaluate predictive performance of a word score, we first measure the
overlap $O(\tau,t)$ between the predicted learning trajectory
$\tau$ and the empirically known trajectory $\tau_{aoa}$ at time $t$ by
counting the number of words that co-occur in $\tau$ and $\tau_{aoa}$ up to
time $t$ and averaging over different $\tau_{aoa}$. We define the word gain $g(\tau,t)$ as the overlap minus random fluctuations, i.e.
\begin{equation}
\centering
g(\tau,t)=O(\tau,t) - \langle O(\tau_{ran},t) \rangle.
\end{equation}

The vocabulary size normalised word gains are computed as $G(\tau,t)=g(\tau,t)/t$ while the word gain Z-Scores consider deviations of overlaps from random guessing in terms of standard deviations $\sigma$, as $Z=g(\tau,t)/\sigma(O(\tau_{ran},t))$.

\subsubsection*{Calculating optimal combinations of layers} 

To allow for varying influences of layers within the multiplex we construct word scores as convex
combinations of individual layer influence, i.e. a word score $s_w$ for
word $w$ is obtained as: 
\begin{equation}
\centering
s_{w}=\alpha s_{w}^{(Asso)}+\beta s_{w}^{(Feat)}+ \gamma s_{w}^{(Co-occ)} + (1-\alpha-\beta-\gamma)s_{w}^{(Phon)},
\end{equation}
where $s_w^{(Asso)}$, $s_w^{(Feat)}$, $s_w^{(Co-occ)}$, and $s_w^{(Phon)}$
are word scores obtained from the respective single layer metrics 
and the coefficients $\alpha$, $\beta$, $\gamma$ give the
influence of each layer on the overall word score $s_w$. This linear combination of
single-layer features is similar in spirit to previous decompositions of
multiplex centrality, see \cite{sola2013eigenvector}
and discussions in \cite{de2015structural}. The optimisation finds the influences $\alpha,\beta,\gamma$ that maximise vocabulary normalised word gain (either over the entire time period or over specific learning phases) after removing 20\% of the words
at random. 20$\%$ was chosen in order not to
remove too many words particularly in the very early word learning stages. Performance is evaluated on all words. Optimisation was performed using a differential evolution method \cite{price2006differential}. Averages are calculated over 50 configurations of left out words for 30 normative age of acquisition orderings. 

\bibliography{biblu3} 

\section*{Author contributions statement}

M.S., N.B.\ and M.B.\ designed the original study, M.S., N.B.\ and M.B.\ conceived the experiments, M.S.\ conducted the experiments, M.S., N.B.\ and M.B.\ analysed the results. All authors reviewed the manuscript. 

\section*{Additional information}

The authors declare no competing financial interests. M.S. was supported by
an EPSRC Doctoral Training Centre grant (EP/G03690X/1). The authors
acknowledge the use of the muxViz software \cite{dedomenico2014muxviz}.

\end{document}